\documentclass[a4paper,twoside,12pt]{article}
\usepackage[dvips]{graphicx}
\usepackage{epsfig}
\newcommand{\eg}{{\em e.g.}}
\newcommand{\ie}{{\em i.e.}}

\newcommand{\als}{$\alpha_s$}

\newcommand{\ccbar}{$c\bar{c}$}

\newcommand{\pythia}{{\sc Pythia}}

\newcommand{\atlas}{{\sc Atlas}}
\newcommand{\lhcb}{{\sc lhc}-b}

\newcommand{\pt}{$p_{\perp}$}
\newcommand{\etal}{{\it et al.}}
\newcommand{\jpsi}{$J/\psi$}
\newcommand{\psip}{${\psi}^{\prime} \,$}
\newcommand{\Y}{$\Upsilon \,$}

\def\lsim{\mathrel{\rlap{\lower4pt\hbox{\hskip1pt$\sim$}}
    \raise1pt\hbox{$<$}}}         
\def\gsim{\mathrel{\rlap{\lower4pt\hbox{\hskip1pt$\sim$}}
    \raise1pt\hbox{$>$}}}         

\begin{document}
\thispagestyle{empty}   
\noindent
TSL/ISV-2001-0243 \\
\vspace*{10mm}
\begin{center}
  \begin{Large}
  \begin{bf}
Prompt J/$\psi$ production at the LHC\\
  \vspace*{5mm}
  \end{bf}
  \end{Large}
  \begin{large}
    J\'er\^ome Damet$^{a}$\footnote{Now at Laboratory for High Energy Physics, University of Bern,
Sidlerstr. 5, CH-3 012 Bern, Switzerland}, Gunnar Ingelman$^{ab}$, Cristiano B. Mariotto$^{ac}$\\  
  \end{large}
\end{center}
$^a$~High Energy Physics, Uppsala University, Box 535, S-75121
Uppsala, Sweden\\
$^b$~Deutsches~Elektronen-Synchrotron~DESY, Notkestrasse~85,~D-22603~Hamburg,~Germany\\
$^c$~Institute of Physics, Univ.\ Fed.\ do Rio Grande do Sul, 
Box 15051, CEP 91501-960 Porto Alegre, Brazil\\
\vspace*{5mm}
\begin{quotation}
\noindent
{\bf Abstract:}
Models with essential non-perturbative QCD dynamics for the production of 
charmonium are extrapolated to give predictions of prompt $J/\psi$ production 
at the LHC. Differences of up to an order of magnitude occurs, although the different models all describe the Tevatron data on high-$p_\perp$ charmonium. An important point is here the treatment of higher order perturbative QCD effects. We also discuss the large rate of prompt $J/\psi$ from these models as a background to CP violation studies. 
\end{quotation}

\section{Introduction}
The important interplay of hard, perturbative QCD (pQCD) and soft, 
non-perturbative QCD effects has recently been clearly demonstrated in the 
production of heavy quarkonia in hadron collisions. The Tevatron data 
\cite{Tevatron} for high-\pt \ \jpsi , \psip  and \Y is up to factors of 50 
above the pQCD prediction in the Colour Singlet Model (CSM) \cite{csm}, where 
a colour singlet \ccbar \ pair is produced at the parton level and gives a 
charmonium state with the same quantum numbers. This deficit can be explained 
by letting a fraction of the more abundant \ccbar \ pairs in a colour octet 
state be transformed to a singlet state through some soft, non-perturbative 
QCD dynamics. Lacking a proper understanding of non-pQCD, this has been 
described in different models: the Colour Octet Model (COM) \cite{com}, 
the Colour Evaporation Model (CEM) \cite{cem}, the Soft Colour Interaction 
model (SCI) \cite{sci} and the Generalized Area Law model (GAL) \cite{gal}. 
All these models can be made to fit these Tevatron data.

In this paper we study the extrapolation of these models to the LHC energy and 
examine the theoretical uncertainty in the charmonium production rate. Future 
LHC data may then discriminate between the models and provide an improved 
understanding of the 
non-pQCD mechanism they involve.

Prompt $J/\psi$ production is also of importance as a background to 
CP-violation studies based on $B$ meson decays into $J/\psi X$, 
with $X$ being $K^0_s$, $\phi$ etc. The $B$ production cross section at the 
LHC is very high in comparison with $e^+e^-$ colliders, but its fraction of 
the total inelastic cross section is small (0.7$\%$). The trigger for $B$ 
physics has therefore to be very selective and is typically based on the 
leptons from a $J/\psi$ decay. Studies of such triggers have shown that 
prompt $J/\psi$ is indeed a source of background \cite{background}. The prompt 
$J/\psi$, which emerge from the primary interaction vertex, can to a large 
extent be distinguished from $J/\psi$ from $B$ decays, which are produced at a 
secondary vertex typically located a few hundred micrometers from the primary 
vertex. There is, however, a remaining prompt charmonium background that 
affects $B$-physics analyses and the associated CP-violation studies, such as 
measurements of $sin2\beta$.

In Section 2 the models and their normalisation to the Tevatron data are 
discussed. The extrapolation to the LHC energy is made in Section 3, where 
also the observed differences are analysed and prompt $J/\psi$ as a background for CP-violation studies is considered. We end with some conclusions in Section 4.

\section{Models and tuning to Tevatron data}\label{tevatron}
In the Colour Octet Model the cross section is factorised in a short distance 
part, where a \ccbar \ pair is produced in a well defined quantum number state 
(${}^{2S+1}L_J$), and a long distance part, giving the probability that this 
state will convert 
non-perturbatively into a charmonium state. These probabilities are given by 
non-relativistic QCD (NRQCD) matrix elements, which  are assumed to be 
universal and in practice are free parameters obtained from fits to 
experimental data. For  high-\pt \ $J/\psi$ production at the Tevatron, the 
main subprocesses are $gg \rightarrow J/\psi g$ and $gq \rightarrow J/\psi q$, which are next-to-leading order in the hard pQCD $c\bar{c}$ production process. The extraction of the NRQCD matrix elements from the Tevatron data has been performed in several steps where perturbative effects and intrinsic transverse momenta have successively been taken better into account explicitly instead of being absorbed into the fitted matrix elements \cite{san1}.

\begin{figure}[thb]
\begin{center}
\mbox{
\epsfig{file=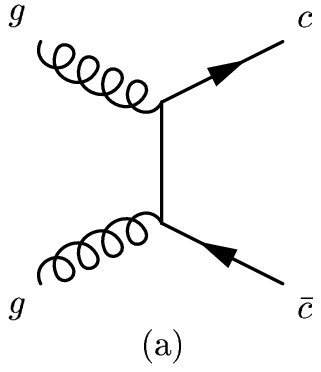,width=4cm}
\hspace{10mm}
\epsfig{file=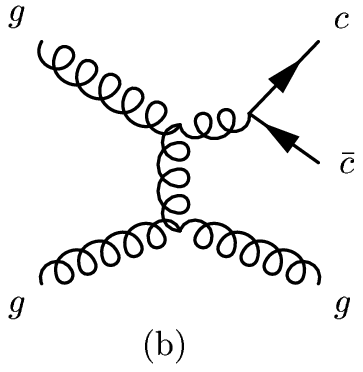,width=4cm}
}
\vspace{0.2cm}
\caption{Illustration of $c\bar{c}$ production processes in (a) leading order ($\alpha_s^2$) and (b) next-to-leading order ($\alpha_s^3$).}
\label{fig:feyn}
\end{center}
\end{figure}

The CEM, SCI and GAL models are based on a similar phenomenological approach, where soft colour interactions can change the colour state of a \ccbar \ pair 
from an octet to a singlet. They employ the same hard pQCD processes to produce a \ccbar \ pair regardless of its spin state. The  leading order (LO) processes are $gg \rightarrow c\bar{c}$ (Fig. \ref{fig:feyn}a) and $q\bar{q} \rightarrow c\bar{c}$. Heavy quark production is, however, known to have large contributions from next-to-leading order (NLO) diagrams \cite{nlo}. Virtual corrections to the leading order processes together with soft and collinear gluon emissions give an increase of the cross section, which can be approximately described as an overall $K$-factor multiplying the leading order cross-section. The LO processes and these NLO processes cannot produce a $J/\psi$ at high-\pt \ since there is nothing to compensate its \pt \ to give the essentially zero \pt \ of the initial partons. Of importance for high-\pt \ $J/\psi$ production is instead NLO tree diagrams with a third hard parton that balances the \pt \ of the $c\bar{c}$ pair. The most important contribution is given by the diagram in Fig.~1b. Although this is an ${\cal O}(\alpha_s)$ correction in terms of a gluon splitting $g\to c\bar{c}$ applied to the basic $2\to 2$ process of $gg\to gg$, it is numerically large since $gg\to gg$ has a much larger ${\cal O}(\alpha_s^2)$ cross section than the LO $c\bar{c}$ production processes.
Matrix elements with non-zero quark masses are only available up to NLO. Still higher orders can be expected to be important at the Tevatron and LHC. The reason is that at these energies many gluons can be emitted and their virtuality need not be very large in order to allow a split into a $c\bar{c}$ pair. 
The higher order processes can be approximately described by the parton shower approach available in the \pythia \ \cite{pythia} Monte Carlo, where in all basic QCD $2\to 2$ processes the incoming and outgoing partons may branch as described by the DGLAP equations \cite{dglap}. 

In CEM the exchange of soft gluons is assumed to give a randomisation of the colour state. This implies a probability $1/9$ that a $c\bar{c}$ pair is in a colour singlet state and produces charmonium if its mass is below the threshold for open charm production, $m_{c\bar{c}}<2m_D$. The fraction of charmonium giving a $J/\psi$ is given by an additional non-perturbative parameter $\rho_{J/\psi}=0.43-0.5$ \cite{cem}. This model was recently implemented in \pythia \ such that 
higher order pQCD processes could be included in terms of parton showers and events be Monte Carlo simulated. It was found \cite{cemmultip} that this model reproduced quite well $x_F$ and $p_\perp$ distributions, both in shape and normalisation, of $J/\psi$ produced in fixed target experiments. 

In the Soft Colour Interaction model \cite{sci} it is assumed that colour-anticolour, corresponding to non-perturbative gluons, can be exchanged between partons emerging from a hard scattering and hadron remnants. This can be viewed as the partons interacting softly with the colour medium, or colour background field, of the initial hadron as they propagate through it. This should be a natural part of the process in which `bare' partons are `dressed' into non-perturbative ones and the confining colour flux tube between them is formed. The hard parton level interactions are given by standard perturbative matrix elements and parton showers, which are not altered by softer non-perturbative effects. The unknown probability to exchange a soft gluon between parton pairs is given by a phenomenological parameter $R$, which is the only free parameter of the model. These colour exchanges lead to different topologies of the confining colour force fields (strings) and thereby to different hadronic final states after hadronisation. This model gives a novel explanation of rapidity gap events in deep inelastic scattering and in hard $p\bar{p}$ processes at the Tevatron \cite{sci,gaps}, which are well reproduced with $R=0.5$. Applying the same Monte Carlo implementation in \pythia \ (with the same $R$-value), it was found that the Tevatron data on high-$p_\perp$ charmonium and bottomonium are also well reproduced \cite{sci-onium}. The increased production rate is here given by the possibility for a perturbatively produced $Q\bar{Q}$ pair in a colour octet state to be transformed to a singlet state as a result of these soft colour interactions. The mapping of $c\bar{c}$ pairs, with mass below the threshold for open charm production, is here made based on spin statistics which avoids  introducing further free parameters. This was also found to give a correct description of the different onium states observed at the Tevatron \cite{sci-onium}. 

An alternative to SCI is the later developed Generalised Area Law model \cite{gal}, where modified colour string-field topologies are obtained by interactions between the strings in the event instead of the partons. A generalisation of the area law suppression $e^{-bA}$, with $A$ the area swept out by the string in energy-momentum space, gives a dynamic probability 
$R=R_0(1-e^{-b\Delta A})$ for two string pieces to interact depending on the area difference $\Delta A$ resulting from the changed string topology. This favours making shorter strings and thereby favours quarkonium production. The parameter $R_0$ and some hadronisation model parameters, \eg \ $b$, are obtained from a fit to data from both deep inelastic scattering and $e^+e^-$ annihilation \cite{gal}.   

\begin{figure}[bht]
\begin{center}
\includegraphics[width=95mm]{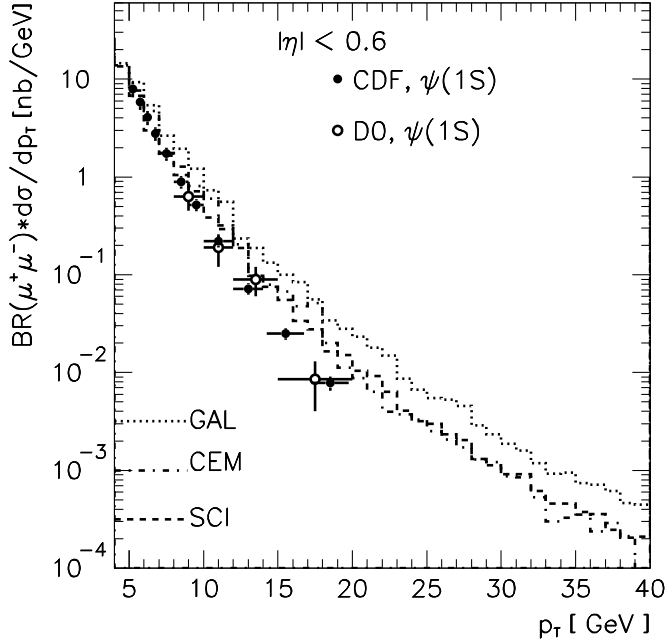}
\vspace*{-0.4cm}
\caption{Distribution in transverse momentum of prompt $J/\psi$ as observed by CDF and D0 \cite{Tevatron} in $p\bar{p}$ interactions at the Tevatron and obtained in the CEM, SCI and GAL models.}
\label{fig:galsci}
\end{center}
\end{figure}

The comparison of the CEM, SCI and GAL models with the Tevatron data is shown 
in Fig.~\ref{fig:galsci}. As can be seen, all models give a quite decent
description of the data. Although the shape is not perfect in the tail of the
distribution, it is quite acceptable given the simplicity of the models. The
overall normalisation is correctly given by the models. For the CEM this is
obtained by setting $\rho_{J/\psi}=0.43$ and the charm quark mass to 1.5 GeV. As
parameterisations of the parton density functions we have used CTEQ4L
\cite{CTEQ4L}, but we have checked that the result does not change much if we
use GRVHO or CTEQ2L (\eg \ CTEQ2L reduces the normalisation by a few percent).
The SCI and GAL models have not been tuned to these data, but the result is also
sensitive to the charm quark mass (taken as default $m_c=1.35$\ GeV in \pythia \ 5.7). The
parton densities are here kept the same as used in other applications of these models, namely CTEQ3L \cite{CTEQ3L} for SCI and CTEQ4L \cite{CTEQ4L} for GAL. The values of the essential parameters $R$ and $R_0$ in SCI and GAL are, as discussed, given by the 
rate of rapidity gap events in deep inelastic scattering. The SCI model is, however, quite stable such that $R$ can be chosen in the range 0.2--0.5. One should note that an arbitrary $K$ factor is not needed in any of the models, since higher order pQCD processes were included through the parton showers. 

In view of the Tevatron running with increased luminosity one can expect 
forthcoming \jpsi \ data extending to higher \pt , which would test and 
constrain the models further. We therefore include in Fig.~\ref{fig:galsci} 
model predictions for \pt \ up to 40~GeV, which show that the CEM and SCI 
models are quite close, whereas the GAL model gives a somewhat higher cross 
section in the high-\pt \ tail. New data could help in discriminating or 
improving the models resulting in reduced uncertainties in the model predictions. 

Based on the ability of these models to reproduce, in a reasonable way, the 
presently available Tevatron data, we now extrapolate them to the LHC energy.

\section{Extrapolation to the LHC energy}\label{lhc}
Applying these models also for the larger energy at the LHC, \ie \ $\sqrt{s}=14$ \ TeV, should be appropriate since they include a reasonable energy dependence. The production of the $c\bar{c}$ pair is given by hard pQCD processes with a well-defined energy dependence. The soft interactions that change colour octet states to singlet states have no explicit energy dependence, similar to the normal hadronisation process. The SCI and GAL models will, however, have an implicit energy dependence since they act on a parton state or string topology that depend on the collision energy. 

\begin{figure}[htb]
\parbox[t]{8cm} {
\epsfxsize=7.5cm
\epsffile{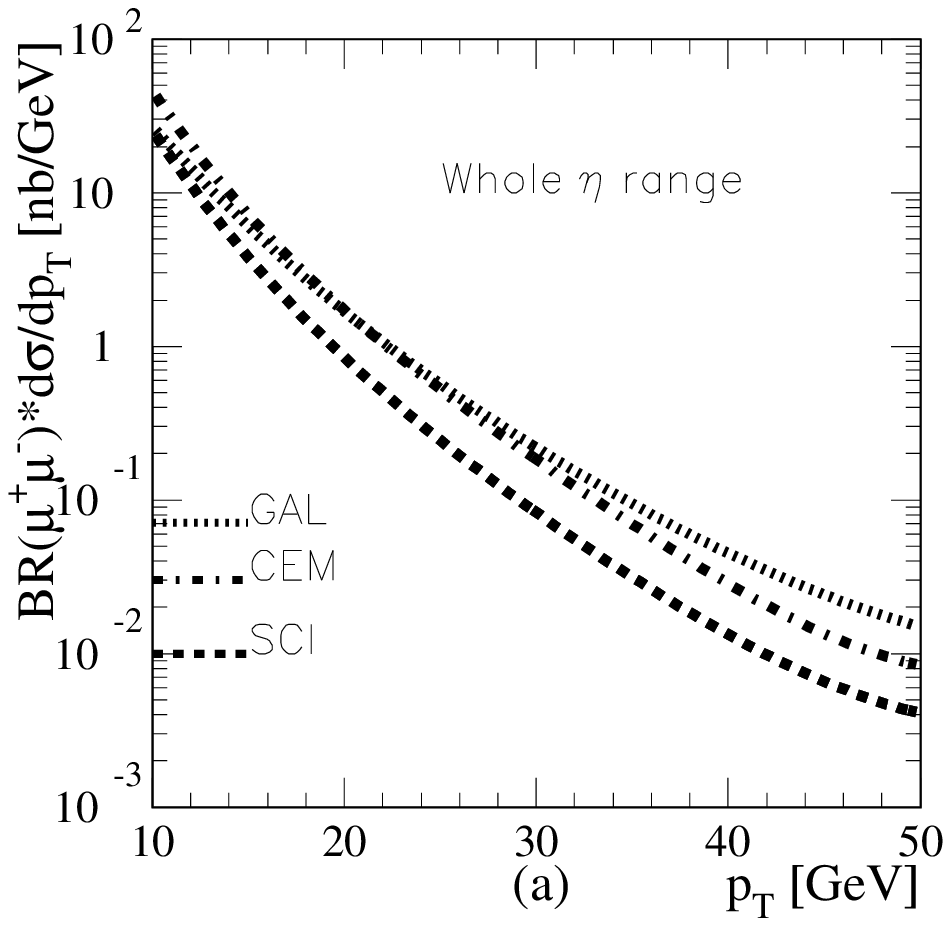}
}
\hspace{0.2cm} \hfil
\parbox[t]{8cm} {
\epsfxsize=7.5cm
\epsffile{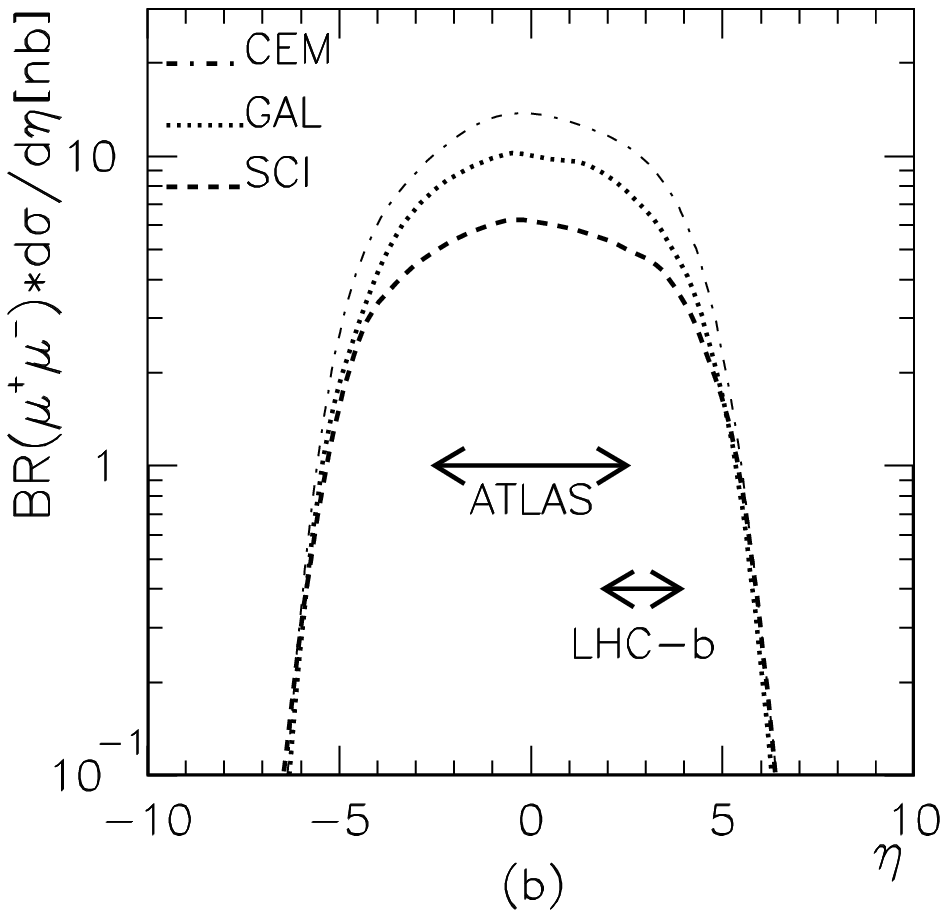}
}

\ \hspace{0.1cm} \hfil \

\parbox[t]{8cm} {
\epsfxsize=7.5cm
\epsffile{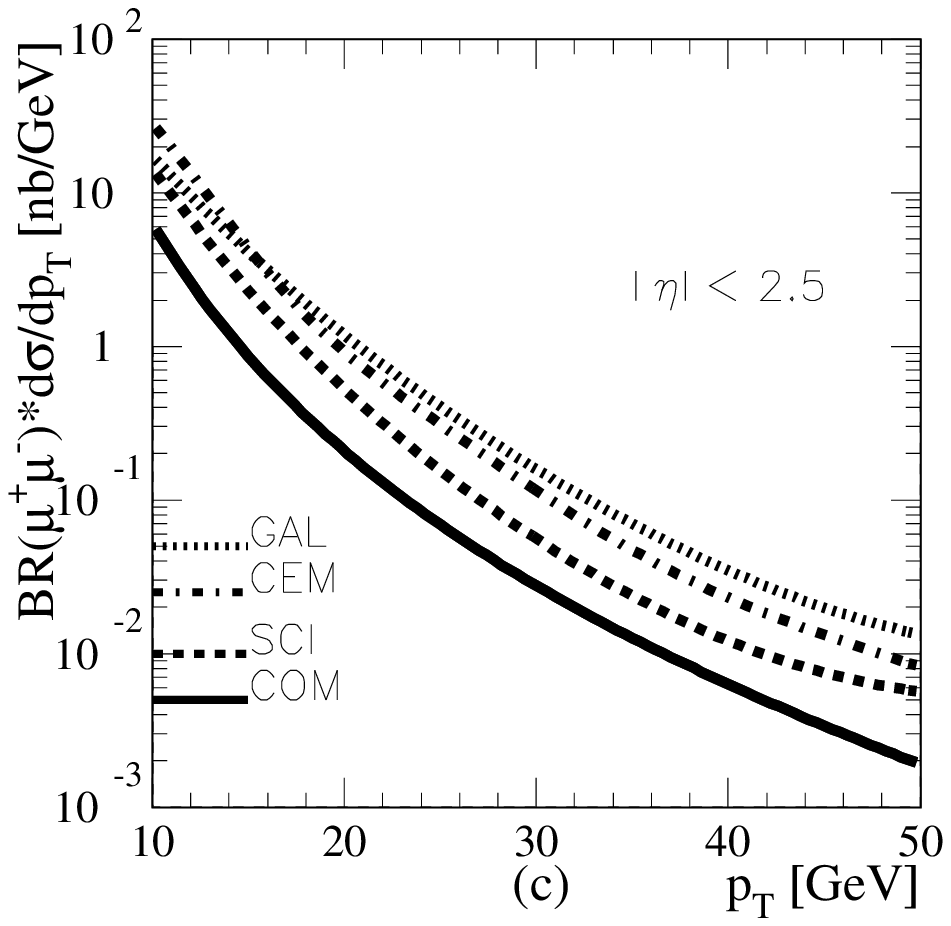}
}
\ \hspace{0.2cm} \hfil \
\parbox[t]{8cm} {
\epsfxsize=7.5cm
\epsffile{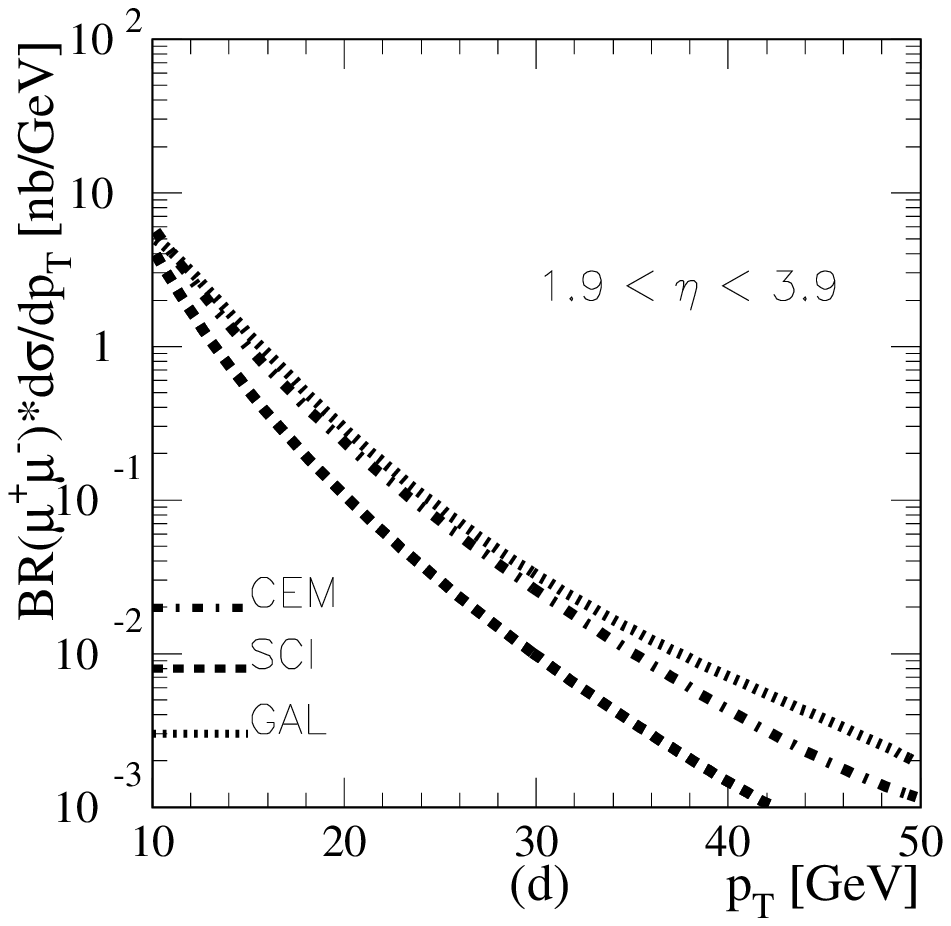}
}
\label{fig:lhc}
\caption{Differential cross sections in transverse momentum and rapidity for 
$J/\psi$ in $pp$ collisions at $\sqrt{s}=14$\ TeV based on the CEM, SCI and GAL
models. In all cases only $J/\psi$ with $p_{\perp}^{J/\psi}>10$\ GeV and decaying into $\mu^+\mu^-$, which in (c, d) are required to be within 
the indicated rapidity coverage of the \atlas \ and \lhcb \ detectors, respectively. For comparison, the COM results from \cite{san1} is included in (c).}
\end{figure}

Results of the models, keeping all parameters fixed from the comparison with the Tevatron data, are shown Fig.~\ref{fig:lhc}. The $p_\perp$ distribution in Fig.~\ref{fig:lhc}a is quite similar for the three models, although they differ somewhat in the high-$p_\perp$ tail and GAL has a somewhat less steep slope. These predictions should, however, not be taken as very precise in view of the simplicity of these models that attempt to describe unknown non-perturbative QCD phenomena. The overall normalisation, which \eg \ is sensitive to the value of the charm quark mass, should not be considered to be better than within about a factor two. 

In order to consider what can realistically be measured, we show in Fig.~\ref{fig:lhc}b 
the distribution in pseudorapidity of $J/\psi$ and the acceptance regions of the LHC experiments. Requiring the muons from the $J/\psi$ decay to be within these regions we obtain the results in Fig.~\ref{fig:lhc}cd, which show lower effective cross sections with slightly more separated model curves. 
Thus, Fig.~\ref{fig:lhc} shows a slightly different behaviour of the models as function of rapidity and transverse momentum, but a more detailed study of this is premature until some data are available to show that the models give a reasonable overall description of \jpsi \ production at the LHC energy. 
A retuning of the models may then reduce these differences and it is therefore 
not clear to what extent these model differences can be exploited to 
discriminate among the models and disentangle details of the charmonium 
production mechanism. In any case, this would require data of quite high 
precision.

Here we can make an interesting comparison with a previous calculation of the 
prompt $J/\psi$ production in the COM model \cite{san1}. As can be seen in 
Fig.~\ref{fig:lhc}c, the COM result is significantly lower than 
the other models. The COM result is based on using pQCD matrix elements in NLO.
At high \pt \ the dominant process is, as discussed above, the order  
${\cal O}(\alpha_s^3)$ process $gg\to c\bar{c}g$ in Fig.~\ref{fig:feyn}b. 
A shift of the momentum distribution of this \ccbar \ pair has been made 
in order to account for the effect of higher order gluon emissions as 
estimated based on the parton shower. Since $g\to c\bar{c}$ in the parton 
shower is here not included, the applied higher order correction does not 
change the normalisation of the \ccbar \ cross section, but only the shape 
of the momentum distribution which becomes somewhat softer due to 
gluon radiation. 

\begin{figure}[htb]
\begin{center}
\mbox{
\epsfig{file=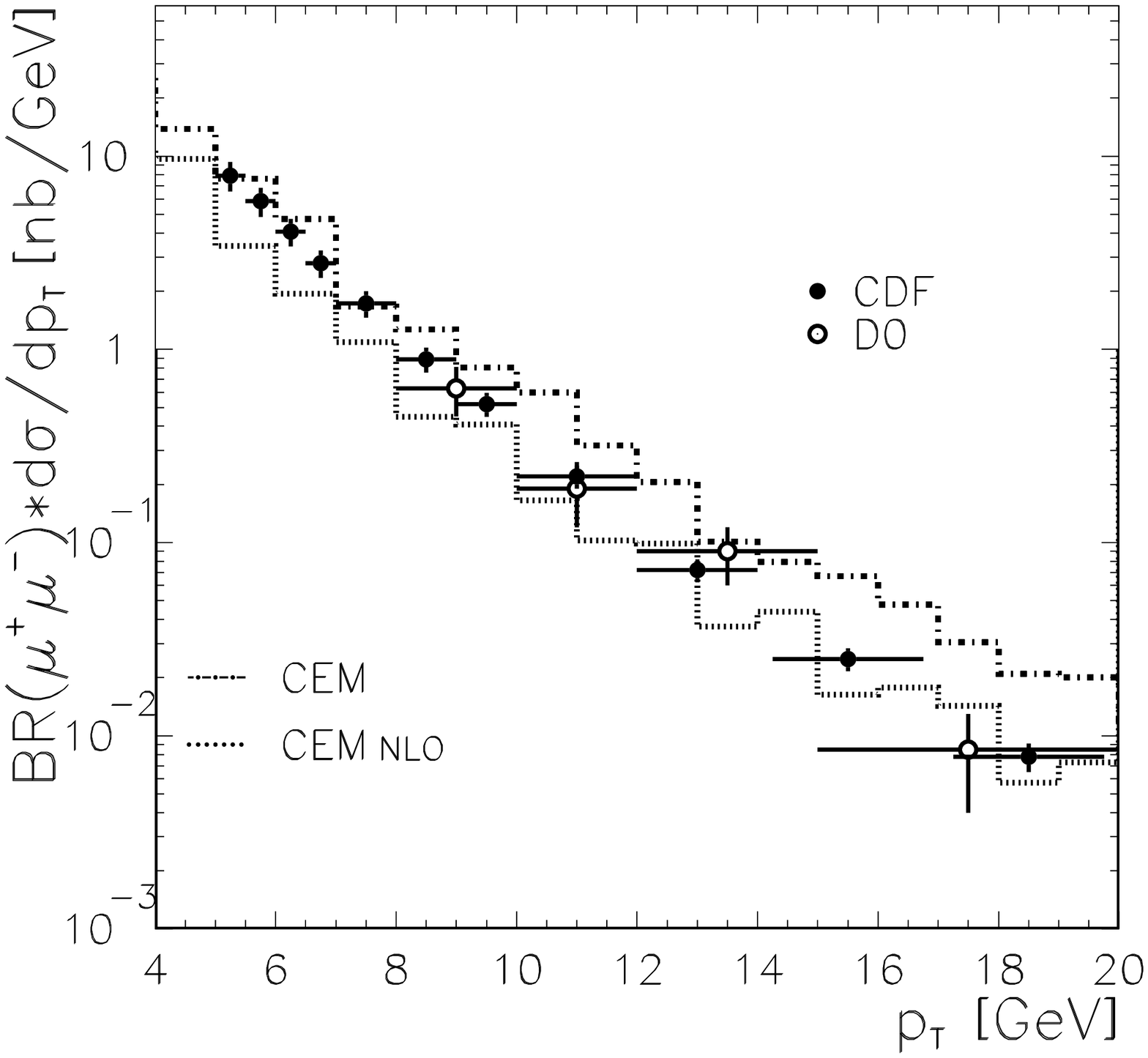,width=8cm}
\epsfig{file=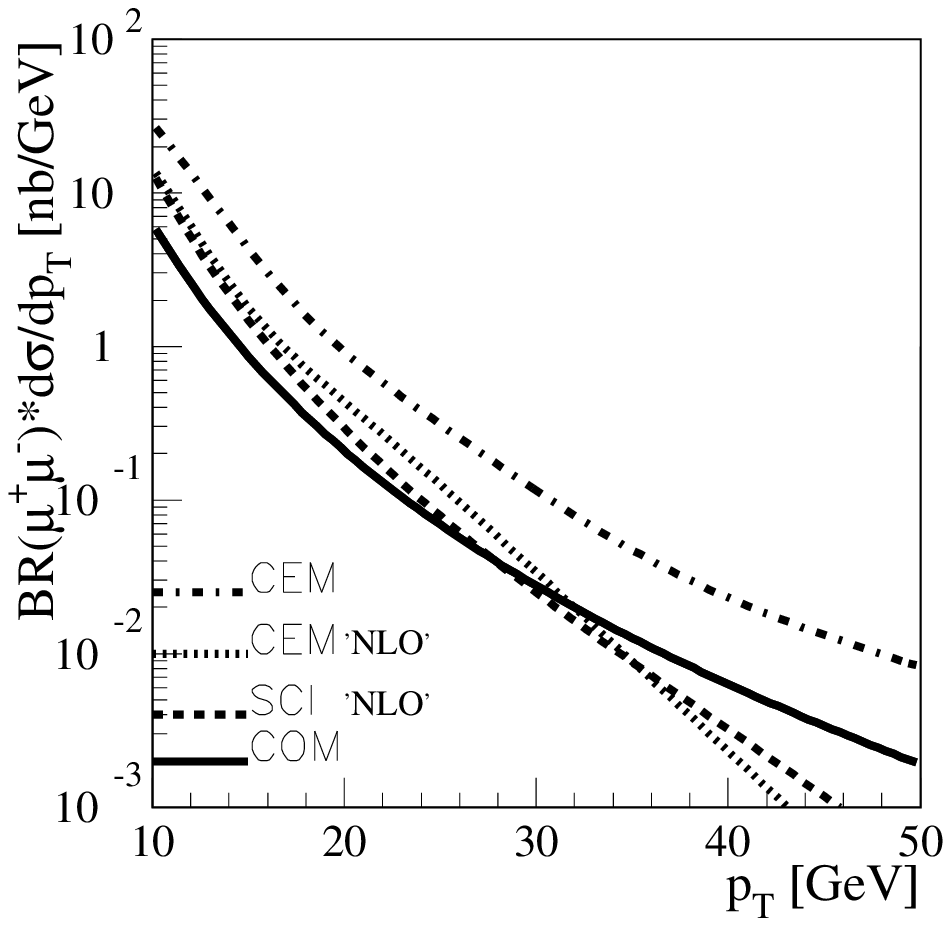,width=8cm}
}
\vspace{0.2cm}
\caption{Distribution in transverse momentum of prompt $J/\psi$ at (a) 
the Tevatron and (b) LHC ($|\eta|<2.5$ as for \atlas ) for the colour 
octet model (COM) based on NLO ${\cal O}(\alpha_s^3)$ matrix elements 
(from \cite{san1}) and for the colour evaporation model (CEM) including 
$g\to c\bar{c}$ to all orders in the parton shower and in the `NLO' 
approximation where only $c\bar{c}$ pairs from the first branching have been 
included. The result of this `first branching' approximation applied 
to the soft colour interaction model (SCI) is also shown in (b).}
\label{gluon}
\end{center}
\end{figure}

Our calculations with the CEM, SCI and GAL models use leading order matrix 
elements and include the full parton shower evolution, including  
$g\to c\bar{c}$ in any branching. This gives an estimate of the \ccbar \ 
production cross section including all higher orders. 
To find out to what extent this is the reason for the observed difference, 
we mimic the ${\cal O}(\alpha_s^3)$ matrix elements by including only those 
$c\bar{c}$ pairs coming from the first branching of the parton shower 
(cf.\ Fig.~1b) giving the results shown in Fig.~\ref{gluon}.
Since we want to explicitly show the effect of omitting higher order \ccbar \ production we have here not changed any parameters in the models (which could be done to partly compensate for the loss of higher orders).  As compared to the standard result, including \ccbar \ from any branching, this `first branching' approximation does indeed give a lower \jpsi \ cross section. 
At the Tevatron energy the difference is not large and could at least partly 
be absorbed into a tuning of parameters such as $m_c$ and \als . 
Nevertheless, in the region of lower \pt \ where the data are more precise, 
some preference for the all order result is indicated. At LHC, 
where more energy is available to build up a more extended parton shower, 
the difference is larger and becomes an order of magnitude at high \pt . 
As can be seen in Fig.~\ref{gluon}b, this reduction of the $J/\psi$ yield 
in the first branching approximation brings the CEM and SCI models closer
to the COM result \cite{san1}. This COM result was based on the older 
parameterisation CTEQ2L of parton densities, but we have checked that this 
causes a reduction in the overall normalisation which is much smaller than 
omitting higher order $g\to c\bar{c}$ in the parton shower. 
Thus, we conclude that $c\bar{c}$ production in orders higher than 
${\cal O}(\alpha_s^3)$ are important at the LHC energy.

As discussed in the Introduction, we also consider the prompt $J/\psi$ production as background for CP-violation studies based on $B$ meson decays 
such as $B_d^0 \rightarrow J/\psi K_s^0$, $B_s^0 \rightarrow J/\psi \phi$, etc. A detailed analysis has been made for \atlas \ \cite{tdr}, taking into account trigger conditions, acceptance cuts, off-line selection criteria and reconstruction methods. This resulted in the signal-to-background ratio of 4.1 for the case of $B_s^0 \rightarrow J/\psi \phi$. 
The estimated contamination from $pp\to J/\psi X$ on the level of 3\% 
was here entirely due to prompt \jpsi \ produced by the COM model implemented 
in the \atlas \ Monte Carlo package. Although other backgrounds were found to be more important, this signal-to-background ratio may be somewhat too optimistic in view of our finding that COM gives a lower prompt \jpsi \ cross section than the CEM, SCI and GAL models. 

At this stage one cannot decide which of these models is most reliable and gives the best prediction for the prompt $J/\psi$ production. The variation between them should therefore be taken as an estimate of the theoretical uncertainty. From Fig.~3c we find cross sections that are almost an order of magnitude larger than in COM. The signal-to-background does not, however, decrease by the full amount of this factor since other backgrounds are also present. 
Nevertheless, this indicates that the prompt $J/\psi$ production could be an important background giving a lower signal-to-background ratio than previously estimated. A proper study of this is beyond the scope of this paper, but our results show that the prompt $J/\psi$ production is a background for CP-violation studies that should not be neglected.

\section{Conclusions}
Charmonium production provides an interesting testing ground for QCD, including both perturbative and non-perturbative effects. Several models (COM, CEM, SCI, GAL) have been developed to account for the possibility that a perturbatively produced $c\bar{c}$ pair in a colour octet state is transformed into a singlet state through soft interactions. This enhances the rate of charmonium production by more than an order of magnitude, as compared to  conventional expectations based on the Colour Singlet Model. All these new models can account for the observed rate of high-$p_\perp$ prompt $J/\psi$ at the Tevatron, and are thereby normalised at this energy. 
Since the models tend to differ more at higher \pt , we have extended our model calculations to give predictions for a region of higher \pt \ which may be reached in the high luminosity runs at the Tevatron. Forthcoming high statistics data may then discriminate among the models or help in reducing the uncertainties in results of the models.

Extrapolating the COM, CEM, SCI and GAL models to the LHC energy, we find significant differences in the predicted prompt $J/\psi$ cross sections; up to almost an order of magnitude. In particular, the COM result is lower than the others. Part of this difference is related to the fact that COM is based on NLO matrix elements, whereas the other models include still higher order $c\bar{c}$ production through the parton shower approximation in pQCD. 

Prompt $J/\psi$ is also a background for studies of $B$ meson decays into $J/\psi$, which are important for studies of CP-violation. An earlier estimate of the signal-to-background ratio based on COM gave the favourable result of 4.1. Using the larger rate of prompt $J/\psi$ from the other models, will reduce the signal-to-background 
ratio. Prompt $J/\psi$ production must therefore be better understood in order to control it as a background for CP-violation studies. 

Improving our understanding of the mechanism for prompt $J/\psi$ production is also important in its own right. It involves an interplay of hard, perturbative and soft, non-perturbative QCD dynamics which is closely connected with the more general problem of understanding QCD. 

{\bf Acknowledgments:} We are grateful to Tord Ekel\"of and Johan Rathsman for 
helpful discussions. This work was supported by the Swedish Natural Science Research Council and the Funda\c{c}\~ao Coordena\c{c}\~ao de 
Aperfei\c{c}oamento de Pessoal de N\'{\i}vel Superior (CAPES), Brazil.

\end{document}